\documentclass[12pt]{article}
\pdfoutput=1
\usepackage{putex}
\usepackage{graphicx}
\usepackage{epstopdf}
\usepackage{enumerate}
\usepackage{cite}
\usepackage{tensor}
\usepackage{slashed}

\newcommand{\abs}[1]{\left\lvert #1 \right\rvert}

\newcommand {\be} {\begin {equation}}
\newcommand {\ee} {\end {equation}}

\newcommand {\bes} {\begin {equation*}}
\newcommand {\ees} {\end {equation*}}

\newcommand{\es}[2] {\begin{equation} \label{#1} \begin{split} #2 \end{split} \end{equation}}

\newcommand{\R}{\mathbb{R}}

\begin{document}

\preprint{PUPT-2377}

\institution{PU}{Joseph Henry Laboratories, Princeton University, Princeton, NJ 08544}
\institution{PCTS}{Center for Theoretical Science, Princeton University, Princeton, NJ 08544}
\institution{IAS}{School of Natural Sciences, Institute for Advanced Study, Princeton, NJ 08540}
\institution{MIT}{Center for Theoretical Physics, Massachusetts Institute of Technology, Cambridge, MA 02139}

\title{$F$-Theorem without Supersymmetry}

\date{}

\authors{Igor R.~Klebanov,\worksat{\PU, \PCTS,\IAS}  Silviu S.~Pufu,\worksat{\PU,\MIT} and Benjamin R.~Safdi\worksat{\PU}}

\abstract{
The conjectured $F$-theorem for three-dimensional field theories states that the finite part of the free energy on $S^3$ decreases along RG trajectories and is stationary at the fixed points.  In previous work various successful tests of this proposal were carried out for theories with ${\cal N}=2$ supersymmetry.  In this paper we perform more general tests that do not rely on supersymmetry. We study perturbatively the RG flows produced by weakly relevant operators and show that the free energy decreases monotonically.  We also consider large $N$ field theories perturbed by relevant double trace operators, free massive field theories, and some Chern-Simons gauge theories. In all cases the free energy in the IR is smaller than in the UV, consistent with the $F$-theorem.  We discuss other odd-dimensional Euclidean theories on $S^d$ and provide evidence that $(-1)^{(d-1)/2} \log |Z|$ decreases along RG flow; in the particular case $d=1$ this is the well-known $g$-theorem.
}

\maketitle


\section{Introduction}

A deep problem in quantum field theory is how to define a measure of the number of degrees of freedom that decreases along any renormalization group (RG) trajectory and is stationary at the RG fixed points. In two-dimensional QFT, an elegant solution to this problem was given by Zamolodchikov \cite{Zamolodchikov:1986gt}, who used the two-point functions of the stress-energy tensor to define a ``$c$-function'' that had the desired properties.   The Zamolodchikov $c$-function has the additional property that at the RG fixed points it coincides with the Weyl anomaly coefficient $c$.  In four-dimensional conformal field theory there are two Weyl anomaly coefficients, $a$ and $c$,  and Cardy has conjectured \cite{Cardy:1988cwa}  that it should be the $a$-coefficient that decreases under RG flow.  This coefficient can be calculated from the expectation of value of the trace of the stress-energy tensor
in the Euclidean theory on $S^4$.

Considerable evidence has been accumulating in favor of the $a$-theorem, especially in supersymmetric 4-d field theories where $a$ is determined by the $U(1)_R$ charges \cite{Anselmi:1997am}. In particular, the principle of $a$-maximization \cite{Intriligator:2003jj}, which states that at superconformal fixed points the correct R-symmetry locally maximizes $a$, has passed many consistency checks that rely both on the field theoretic methods and on the AdS/CFT correspondence \cite{Maldacena:1997re,Gubser:1998bc,Witten:1998qj}. For large $N$ superconformal gauge theories dual to type IIB string theory on $AdS_5\times Y_5$, $Y_5$ being a Sasaki-Einstein space, $a$-maximization is equivalent to the statement \cite{Martelli:2005tp} that the Sasaki-Einstein metric on $Y_5$ is a volume minimizer within the set of all Sasakian metrics on this space.  This equivalence was proved in \cite{Butti:2005vn,Eager:2010yu}.  Very recently, a general proof of the $a$-theorem was constructed in \cite{Komargodski:2011vj}.

Due to the abundance of fixed points in three-dimensional QFT, as well as their relevance to observable phase transitions, it is of obvious interest to find a 3-d analogue of the 2-d $c$-theorem and of the 4-d $a$-theorem. Such a result would establish very general restrictions on RG flows. An obvious difficulty, however, is that because of the absence of a conformal anomaly in odd dimensions, the trace of the stress-energy tensor
simply vanishes at the RG fixed points and hence cannot be a measure of the number of degrees of freedom. A physically reasonable measure could be the free energy at finite temperature $T$ \cite{Appelquist:1999hr}. At RG fixed points, this quantity can be extracted from the Euclidean theory on $\R^{d-1}\times S^1$:
 \es{Fthermal}{F_T = -{\Gamma(d/2)\zeta(d/2)\over \pi^{d/2}} c_{\text{Therm}} V_{d-1} T^d \, ,
 }
 where $V_{d-1}$ is the spatial volume, and $c_{\text{Therm}}$ is a dimensionless number normalized so that a massless scalar field gives $c_{\text{Therm}}=1$.  However, there are cases in dimensions $d>2$ where $c_{\text{Therm}}$ increases along RG flow:  such a behavior occurs, for example, in the flow from the critical $d=3$ $O(N)$ model to the Goldstone phase described by $N-1$ free fields \cite{Sachdev:1993pr,Chubukov:1994zz}.\footnote{We thank Subir Sachdev for telling us about this example.}  This rules out the possibility of a $c_{\text{Therm}}$-theorem and therefore of $c_{\text{Therm}}$ being a `good' measure of the number of degrees of freedom.\footnote{Another reason why there cannot be a $c_{\text{Therm}}$-theorem is that $c_{\text{Therm}}$ varies along lines of fixed points, as is well-known for example in the four-dimensional ${\cal N}=4$ supersymmetric Yang-Mills theory \cite{Gubser:1998nz}.
}
Another proposal, which associates a measure of the degrees of freedom in $d=3$ with the coefficient $c_T$ of the correlation function of two stress-energy tensors, was made in \cite{Petkou:1995vu, Barnes:2005bm}.

Quite recently, a new proposal was made for a good measure of the number of degrees of freedom in a 3-d Euclidean CFT \cite{Jafferis:2010un,Jafferis:2011zi}:
\es{MtheoryExpectation}{
  F = -\log |Z_{S^3}| \,,
 }
where $Z_{S^3}$ is the Euclidean path integral of the CFT conformally mapped to $S^3$.  Jafferis \cite{Jafferis:2010un} conjectured that the 3-d analogue of $a$-maximization is that the R-symmetry of
${\cal N}=2$ superconformal theories in three dimensions extremizes $F$. In \cite{Jafferis:2011zi} it was further conjectured that in unitary 3-d CFT $F$ is positive and that it decreases along any RG flow;\footnote{The three-sphere free energy is ambiguous along RG flow, but is well-defined for any conformal field theory. Thus, a precise statement of the conjecture is that, if there is smooth RG flow from CFT$_1$ to CFT$_2$ then their three-sphere free energies satisfy $F_1 > F_2$.} various ${\cal N}=2$ supersymmetric examples were presented in support of these statements.
These conjectures were preceded by the important work \cite{Pestun:2007rz,Kapustin:2009kz} on path integrals in supersymmetric gauge theories on spheres, where these infinite-dimensional path integrals were reduced using the method of localization \cite{Nekrasov:2002qd} to certain finite-dimensional matrix integrals. The solution of these matrix models in the large $N$ limit \cite{Drukker:2010nc, Herzog:2010hf, Santamaria:2010dm, Martelli:2011qj, Cheon:2011vi, Jafferis:2011zi} produced perfect agreement
with the $AdS_4/CFT_3$ correspondence, explaining in particular the $N^{3/2}$ scaling of the number of degrees of freedom expected on the gravity side of the duality \cite{Klebanov:1996un}.  By now there exists considerable evidence that in ${\cal N}=2$ supersymmetric 3-d field theory the R-symmetry maximizes $F$ at the fixed points
and that $F$ decreases under RG flows. These ideas have passed some field theoretic tests \cite{Jafferis:2010un, Amariti:2011hw, Jafferis:2011ns,Niarchos:2011sn,Willett:2011gp,Minwalla:2011ma,Amariti:2011da, Benvenuti:2011ga, Gulotta:2011si,Amariti:2011xp, Nishioka:2011dq}, and various issues in defining supersymmetric theories on $S^3$ were clarified in \cite{Festuccia:2011ws}. Also, for large $N$ theories with $AdS_4\times Y_7$ dual descriptions in M-theory, $F$-maximization is correctly mapped to the minimization of the volume of the Sasaki-Einstein spaces $Y_7$ \cite{Martelli:2011qj,Jafferis:2011zi}.

Other ideas on how to best define the number of degrees of freedom in field theory, this time in Minkowski signature, were advanced in \cite{Myers:2010tj}.  For a field theory on $\R^{d-1,1}$, it was proposed in \cite{Myers:2010tj} that one can consider the entanglement entropy across an $S^{d-2}$ boundary. It was further shown in \cite{Casini:2011kv} (see also \cite{Dowker:2010yj}) that if the field theory in question is a CFT, this entanglement entropy agrees with the free energy of the Euclidean CFT on $S^d$.

In this paper we will subject the $F$-theorem to additional tests that do not rely on supersymmetry.  Our approach is similar to Cardy's \cite{Cardy:1988cwa} who, in the absence of a proof of the $a$-theorem,
presented some evidence for it in the context of a CFT on $S^4$ perturbed by weakly relevant operators.  (His work generalizes similar calculations in $d=2$ \cite{Zamolodchikov:1986gt,Ludwig:1987gs}.)  In sections 2 and 3 we use the perturbed conformal field theory on $S^3$ to present evidence for the $F$-theorem. We also discuss other odd-dimensional Euclidean theories on $S^d$ where similar perturbative calculations provide evidence that $(-1)^{(d-1)/2} \log |Z|$ decreases along RG flow. In the particular case $d=1$ these calculations were carried out in
\cite{Affleck:1991tk,Affleck:1992ng} providing evidence for the $g$-theorem.\footnote{The $d=1$ dynamics is often assumed to take place on the boundary of a $d=2$ conformal field theory. A proof of the $g$-theorem in this context was given in \cite{Friedan:2003yc}.}
In section 4 we review the calculations of $F$ for theories involving free massless boson, fermion, and vector fields.
We show that these values are consistent with the $F$-theorem for some RG flows.  In section 5 we consider another class of examples which
involve RG flows in large $N$ field theories perturbed by relevant double-trace operators.  In these cases, the theories flow to IR fixed points, and $F_\text{IR}- F_\text{UV}$ can be calculated even when the double-trace operator is not weakly relevant \cite{Gubser:2002vv,Diaz:2007an,Allais:2010qq}. The results are consistent with the $F$-theorem.  An explicit example of this kind is the critical $O(N)$ model. In particular, we show that the flow from the critical $O(N)$ model to the Goldstone phase, which was earlier found to violate the $c_{\text{Therm}}$-theorem \cite{Sachdev:1993pr,Chubukov:1994zz}, does not violate the $F$-theorem.

\section{Perturbed Conformal Field Theory}

\label{pert}

In this section we discuss Euclidean conformal field theories perturbed by a slightly relevant scalar operator of dimension $\Delta = d-\epsilon$, where $0< \epsilon \ll 1$. Our approach follows closely that in
\cite{Zamolodchikov:1986gt,Ludwig:1987gs,Cardy:1988cwa,Affleck:1991tk,Affleck:1992ng}.
To keep the discussion fairly general, we will work in an arbitrary odd dimension $d$ throughout most of the following calculation,
 though the case of most interest to us is $d=3$. Our calculations generalize those carried out for $d=1$ to provide evidence for the $g$-theorem \cite{Affleck:1991tk,Affleck:1992ng,Callan:1995em}.  We take the action of the perturbed field theory to be
\es{pertaction}{
S = S_0 + \lambda_0 \int d^d x \sqrt{G} O(x) \, ,
}
where $S_0$ is the action of the field theory at the UV fixed point, $\lambda_0$ is the UV bare coupling defined at some UV scale $\mu_0$, $G$ is the determinant of the background metric, and $O(x)$ the bare operator of dimension $\Delta$.

\subsection{Beta function and the running coupling}

For the purposes of finding the beta function it is sufficient to work in the flat $\R^d$.
For the CFT on $\R^d$, conformal invariance fixes the functional form of the connected two-point and three-point functions \cite{Polyakov:1970xd}, and we choose the normalization of $O$ to be such that
 \es{CorrelatorsRd}{
  \langle O(x) O(y) \rangle_0 &= \frac{1}{\abs{x-y}^{2 (d-\epsilon)}} \,, \\
   \langle O(x) O(y) O(z) \rangle_0 &= \frac{C}{\abs{x-y}^{d-\epsilon} \abs{y-z}^{d-\epsilon} \abs{z-x}^{d-\epsilon}} \,,
 }
for some constant $C$.  These correlators correspond to the OPE
  \es{OPE}{
   O(x) O(y) = \frac{1}{\abs{x-y}^{2(d-\epsilon)}} + \frac{C O(x)}{\abs{x-y}^{d-\epsilon}} + \ldots \qquad \text{as $x \to y$} \,.
  }

In the perturbed theory, the coupling runs.  The beta function is \cite{Polyakov:1987ez,Cardy:1988cwa}\footnote{This equation differs from eq.~(9) in \cite{Cardy:1988cwa} by the sign of the second term because our coupling $g$ differs from the one in \cite{Cardy:1988cwa} by a minus sign.}
\es{betafun}{
  \beta(g)  = \mu \frac{d g}{d \mu } = - \epsilon g + \frac{\pi^{d/2}}{\Gamma\left({d\over2}\right)} C g^2 + {\cal O}(g^3)\ ,
  }
where $\mu$ is the renormalization scale, and $g =\lambda \mu^{-\epsilon}$ is the dimensionless renormalized coupling.  Integrating this equation with the boundary condition $g(\mu_0) = g_0\ll 1$, where $\mu_0$ is a UV cutoff, we obtain the running coupling
 \es{gIntegrated}{
  g(\mu) = g_0 \left( \frac{\mu_0}{\mu} \right)^\epsilon - \frac{\pi^{d/2}}{\Gamma\left({d\over 2}\right)} \frac{C g_0^2}{2 \epsilon} \left[ \left(\frac{\mu_0}{\mu} \right)^{2 \epsilon} - \left(\frac{\mu_0}{\mu} \right)^\epsilon \right] + {\cal O}(g_0^3)\,.
 }
One can understand the two equations above from the following RG argument.
Correlation functions in the interacting theory differ from the ones in the free theory by an extra insertion of
 \es{ExponentialExpanded}{
   e^{-\lambda_0 \int d^d x\, O(x)} &= 1 -\lambda_0 \int d^d x\, O(x) + \frac{\lambda_0^2}{2} \int d^d x\, \int_{\abs{x-y}>\frac{1}{\mu_0}} d^d y\, O(x) O(y) + \ldots \,,
 }
where the condition $\abs{x-y}>\frac{1}{\mu_0}$ comes from imposing the UV cutoff $\mu_0$.  In obtaining an effective action at some scale $\mu$, one simply isolates the contribution from modes between energy scales $\mu_0$ and $\mu$: for example, we write the last integral in \eqref{ExponentialExpanded} as
 \es{IntegrateOut}{
   \int_{\abs{x-y}>\frac{1}{\mu_0}} d^d y\, O(x) O(y) =
    \int_{\abs{x-y}>\frac{1}{\mu}} d^d y\, O(x) O(y) + \int_{\frac{1}{\mu_0}<\abs{x-y}<\frac{1}{\mu}}
      \frac{d^d y\, C O(x)}{\abs{x-y}^{d-\epsilon}} + \ldots
 }
where in the region $\frac{1}{\mu_0}<\abs{x-y}<\frac{1}{\mu}$ we only exhibited the contribution from the second term in the OPE \eqref{OPE}.  The first term in eq.~\eqref{IntegrateOut} should be thought of as arising from the effective action at scale $\mu$, while the second term should be interpreted as a renormalization of the coupling.  Combining \eqref{IntegrateOut} with \eqref{ExponentialExpanded}, one can deduce that the effective coupling $\lambda(\mu)$ is:
 \es{lambdaRunning}{
  \lambda(\mu) = \lambda_0 - \frac{C \lambda_0^2}{2} \int_{\frac{1}{\mu_0}< \abs{x-y}<\frac{1}{\mu}} \frac{d^d y}{\abs{x-y}^{d-\epsilon}} + \ldots = \lambda_0 - \frac{C \lambda_0^2}{2\epsilon} \Vol(S^{d-1}) \left[\frac{1}{\mu^\epsilon} - \frac{1}{\mu_0^\epsilon} \right] \,.
 }
Using $\Vol(S^{d-1}) = 2 \pi^{d/2} / \Gamma(d/2)$, one can further check that this expression agrees with \eqref{gIntegrated} provided $g_0 = \lambda_0 \mu_0^{-\epsilon}$ and $g(\mu) = \lambda(\mu) \mu^{-\epsilon}$.

If $C<0$ then both terms in the beta function \eqref{betafun} are positive; thus, $g$ grows along the flow, and the fate of the IR theory depends on the coefficients of the
$g^3$ and higher order terms.  However, if $C>0$ then there exists a robust IR fixed point at
  \es{gstar}{
  g^* =  \frac{\Gamma\left({d \over2}\right) \epsilon}{\pi^{d/2} C}+ {\cal O}(\epsilon^2) \, , }
  whose position depends on the coefficient of the $g^3$ term only through the terms of order $\epsilon^2$.

\subsection{Free energy on $S^d$}

From now on, let us consider the field theory on a $d$-dimensional sphere $S^d$ of radius $a$.  Putting the field theory on $S^d$ effectively sets the RG scale $\mu$ to be of order $1/a$.  For convenience we will set $\mu = 1/(2a)$ and express our answers in terms of the renormalized coupling $g$ at this scale.  In the following computations we will also send the UV cutoff $\mu_0 \to \infty$ after appropriately subtracting any UV divergences.

The metric on $S^d$ is most conveniently described through stereographic projection to $\R^d$ because in these coordinates the metric is manifestly conformally flat:
 \es{Stereographic}{
  ds^2 = \frac{4 a^2}{\left(1 + \abs{x}^2\right)^2} \sum_{i = 1}^d (dx_i)^2 \,, \qquad
   \abs{x}^2 \equiv \sum_{i = 1}^d (x_i)^2 \,.
 }
In the unperturbed theory, the connected correlation functions of $O$ on $S^d$ can be obtained from those in flat space given in eq.~\eqref{CorrelatorsRd} by conformal transformation:
 \es{npnt}{
  \left< O(x) O(y) \right>_0 &= \frac{1}{s(x, y)^{2(d-\epsilon)}} \, , \\
  \left<O(x) O(y) O(z) \right>_0 &=\frac{C}{s(x,y)^{d-\epsilon} s(y,z)^{d-\epsilon} s(z,x)^{d-\epsilon}} \, ,
  }
where
  \es{sDef}{
   s(x, y) = 2 a \frac{\abs{x-y}}{\left(1 + \abs{x}^2\right)^{1/2} \left(1 + \abs{y}^2\right)^{1/2}}
  }
is the ``chordal distance'' between points $x$ and $y$.

The path integral on $S^3$ has UV divergences that should be subtracted away.  After this regularization, which we will perform through analytic continuation, one can essentially remove the UV cutoff $\mu_0$ by sending it to infinity.  The resulting regularized path integral $Z_0(\lambda_0)$ depends on the bare coupling $\lambda_0$.  As is standard in perturbative field theory, one can write down the following series expansion for $\log| Z(\lambda_0)|$ in terms of the connected correlators of the unperturbed theory:
 \es{ZExpansion}{
   \log \left| \frac{Z(\lambda_0)}{Z(0)} \right|
   = \sum_{n=1}^\infty \frac{(-\lambda_0)^n}{n!} \int d^d x_1\sqrt{G} \dotsb \int d^d x_n \sqrt{G} \langle O(x_1) \dotsb O(x_n) \rangle_0 \,.
 }
We have $\langle O(x) \rangle_0 = 0$ because the unperturbed theory is a CFT.  Using the definition $F \equiv - \log Z$, we can write the first few terms in the above expression as
\es{ffcn}{
\delta F(\lambda_0) \equiv F(\lambda_0) - F(0)= - \frac{\lambda_0^2}{2} I_2 + \frac{\lambda_0^3}{6} I_3 + {\cal O}(\lambda_0^4) \,,
}
where
  \es{results}{
 I_2 = \int d^dx \sqrt{G}  \int d^dy \sqrt{G}  \left< O(x) O(y) \right>_0 &= \frac{ (2a)^{2\epsilon} \pi^{d+ 1/2}}{2^{d-1}} \frac{ \Gamma\left( -{d\over2} +\epsilon \right)}{\Gamma\left( d+1\over 2\right) \Gamma(\epsilon)} \,,\\
 I_3 = \int d^dx \sqrt{G} \int d^dy \sqrt{G}  \int d^dz \sqrt{G} \left< O(x) O(y) O(z) \right>_0 &= \frac{ 8 \pi^{3 (d+1)/2}  a^{3 \epsilon}}{\Gamma(d)} \frac{ \Gamma \left( -{d\over2} + {3 \epsilon \over 2} \right)}{\Gamma \left( 1+\epsilon \over 2\right)^3} C \,.
  }
These integrals were evaluated through analytic continuation in $\epsilon$ from a region where they are absolutely convergent \cite{Cardy:1988cwa}.

One can simplify equation~\eqref{ffcn} by expressing it in terms of the renormalized coupling $g$ instead of the bare coupling $\lambda_0$ and performing a series expansion in $\epsilon$.  Solving eq.~\eqref{lambdaRunning} for $\lambda_0$ (with $\mu_0 \to \infty$ and $\mu = 1/(2a)$), one obtains
\es{lambdageqn}{
\lambda_0 (2a)^{\epsilon} = g + \frac{C \pi^{d/2}}{\epsilon \Gamma\left(\frac d2 \right)} g^2 + {\cal O}(g^3) \,.
}
Substituting this expression together with the expressions for $I_2$ and $I_3$ from eq.~\eqref{results} into eq.~\eqref{ffcn} gives in odd dimensions $d$
\es{ffcn1}{
\delta F(g) = (-1)^{\frac{d+1}{2}} \frac{2 \pi^{d+1}}{d!} \left[- \frac{1}{2} \epsilon g^2 +\frac{1}{3} \frac{\pi^{d/2}}{\Gamma\left(\frac{d}{2} \right)} C g^3 + {\cal O}(g^4) \right] \,,
}
where we expanded each coefficient of $g^n$ to the first nonvanishing order in $\epsilon$.  By comparing this formula with the beta function \eqref{betafun}, we observe that, to the third order in $g$, the derivative of the free energy is proportional to the beta function:
 \es{DerivBeta}{
  \frac{ d F}{d g} = (-1)^{\frac{d+1}{2}}\frac{2 \pi^{d+1}}{d!} \beta(g) + {\cal O}(g^2) \,.
 }
The proportionality between $dF/dg$ and $\beta(g)$ to this order in perturbation theory is not unexpected:  one can show that $ dF(g)/dg$ equals the integrated one-point function of the renormalized operator $O$,
\es{dfdg}{
\frac{d F}{d g} &=    \mu^{\epsilon} \int d^dx \sqrt{G}   \langle O_\text{ren}(x) \rangle_\lambda \,.
}
This one point function is required by conformal invariance to vanish at the RG fixed points.  To the order in $g$ we have been working at, both the beta function and the one point function of $O$ are quadratic functions, so the fact that conformal invariance forces them to have the same zeroes implies that they must be proportional.  To higher orders in perturbation theory, we expect that $dF/dg$ will equal $\beta(g)$ times a nonvanishing function of $g$.

One can also note that for both signs of $C$ the beta function $\beta(g)$ is negative to second order in perturbation theory.  Eq.~\eqref{DerivBeta} then tells us that the quantity $\tilde F = (-1)^{\frac{d+1}{2}} F$ is a monotonically decreasing function of the radius of the sphere in all odd dimensions.  We interpret this behavior as a monotonic decrease in $\tilde F$ along RG flow between the UV and IR fixed points.  $\tilde F$ is stationary at conformal fixed points, supporting the $F$-theorem in three dimensions and the $g$-theorem in one dimension.

Recall that when $C>0$ there is a perturbative fixed point at the value of the coupling $g^*$ given in \eqref{gstar}.  Eq.~\eqref{ffcn1} tells us that the difference between the free energy at this perturbative fixed point and that at the UV fixed point $g=0$ is
\es{deltafdiffgen}{
\delta \tilde F(g^*) = - \frac{ (d-2)!! }{ 2^{d-1} d (d-1)!!}\frac{\pi^2 \epsilon^3}{3 C^2} \, .
}
The case of most interest is $d=3$, where
\es{deltafdiff}{
\left. \delta F(g^*) \right|_{d=3}  = - \frac{\pi^2 \epsilon^3}{ 72 C^2} \, .
}
We will be able to reproduce this expression in a specific example in section~\ref{bosdeff}.

The arguments above relied heavily on $O(x)$ being a scalar operator.  If instead $O(x)$ is a pseudo-scalar, then the relation
\es{psuedo}{
\langle O(-x_1) O(-x_2) \ldots O(-x_n) \rangle_0  = (-1)^n \langle O(x_1) O(x_2)\ldots O(x_n) \rangle_0 \, .
}
implies that the integrated $n$-point functions of $O(x)$ vanish if $n$ is odd.  In particular $I_3 = 0$ in equation~\eqref{results}, and so the first non-linear correction to the beta function is of order $g^3$; it comes from integrating the four-point function of $O(x)$ as opposed to the three-point function as was the case for a scalar operator.  Because the form of the four-point function is not fixed by conformal invariance but rather depends on the details of the theory, it is hard to say anything general in this case.  A specific example of a slightly relevant pseudo-scalar deformation is discussed in section~\ref{fermdouble}, with the deformation coming from a fermionic double trace operator.

\section{Towards a more general proof of the F-theorem}

Let us consider a CFT on $S^d$ perturbed by multiple operators,
\es{actionGEN}{
S = S_0 + \lambda^i_0 \int d^dx \sqrt{G} O_{i}(x) \, ,
}
where the bare operators $O_{i}$ have dimensions $\Delta_i = d - \epsilon_i$ with $\epsilon_i>0$, and $S_0$ is a conformally-invariant action.  In section~\ref{pert} we studied the special case where there was only one such perturbing operator.

In terms of the dimensionless running couplings $g^i$, which we will denote collectively by ${\bf g}$, a simple application of the chain rule gives
\es{dfdmGEN} {
\frac{ d F}{d \log \mu} = \beta^i({\bf g})  \frac{ \partial F}{\partial g^i} \,, \qquad
 \beta^i({\bf g}) \equiv  \frac{d g^i}{d \log \mu} \,,
}
where we introduced the beta functions $\beta^i({\bf g})$.  Differentiating the partition function with respect to $g^i$, one can see that the gradients $\partial F / \partial g^i$ are given by the general relation:
\es{oneptGEN}{
 \frac{ \partial F}{\partial g^i} &= \mu^{\epsilon_i} \int d^dx \sqrt{G} \left< O_{\text{ren}i}(x) \right>_\lambda \\
 &= (-1)^{\frac{d+1}{2}}\frac{2 \pi^{d+1}}{d!}  h_{i j} (\bold{g}) \beta^j \, .
 }
In the last line of this eq.~\eqref{oneptGEN} we have defined the matrix $h_{i j} ({\bf g})$, which can be thought of as a metric on the space of coupling constants. Consequently, introducing $\tilde F = (-1)^{\frac{d+1}{2}} F$ as in the previous section, we have\footnote{This equation is analogous to that derived for the $c$-function in two dimensional field theory \cite{Zamolodchikov:1986gt}, where it contains a metric on the space of coupling constants well-known as the Zamolodchikov metric.}
\es{dfdmGENnew}{
\frac{ d \tilde{ F}}{d \log{\mu}} = \beta^i h_{ij} \beta^j \,.
}
In principle, the entries of the matrix $h_{ij}({\bf g})$ could be singular for certain values of the coupling.  A sufficient condition for the $F$-theorem to hold is that $h_{ij}({\bf g})$ is  strictly positive definite for all ${\bf g}$.  We will see that this is the case at least perturbatively in small ${\bf g}$.

The perturbative construction of $\beta^i (\bold{g})$ and $h_{ij} (\bold{g})$ generalizes the computation in section~\ref{pert}.  We can choose our operators $O_i(x)$ so that in flat space the two and three-point functions at the UV fixed point are
 \es{CorrelatorsRdGEN}{
  \langle O_{i} (x) O_{j} (y) \rangle_0 &=  \frac{\delta_{ij}}{\abs{x-y}^{2 \Delta_i}} \,, \\
   \langle O_{i}(x) O_{j}(y) O_{k}(z) \rangle_0 &= \frac{C_{ijk}}{\abs{x-y}^{\Delta_i+ \Delta_j - \Delta_k} \abs{y-z}^{\Delta_j + \Delta_k- \Delta_i } \abs{z-x}^{\Delta_i + \Delta_k - \Delta_j}} \,,
 }
for some structure constants $C_{ijk}$.
The corresponding OPE is
  \es{OPEMultiple}{
   O_i(x) O_j(y) = \frac{\delta_{ij}}{\abs{x-y}^{2 \Delta_i}} + \frac{C^k_{ij} O_k(x)}{\abs{x-y}^{\Delta_i+ \Delta_j - \Delta_k}} + \ldots \qquad \text{as $x \to y$} \,,
  }
where $C^k_{ij}=\delta^{kl} C_{lij}$.  These correlators yield the beta functions
\es{betafunManyOps}{
  \beta^i({\bf g})  = \mu \frac{d g^i}{d \mu } = - \epsilon_i g^i + \frac{\pi^{d/2}}{\Gamma\left({d\over2}\right)} \sum_{j,k} C^i_{jk} g^j g^k + {\cal O}({\bf g}^3)\ ,
  }
  and the free energy
\es{Freegen} {
  \delta F= (-1)^{\frac{d+1}{2}}\frac{2 \pi^{d+1}}{d!} \left [- \frac{1}{2} \sum_i \epsilon_i (g^i)^2 + \frac{\pi^{d/2}}{3\Gamma\left({d\over2}\right)} \sum_{i,j,k} C_{ijk} g^i g^j g^k + {\cal O}({\bf g}^4) \right ]\ .
  }
We see that (\ref{oneptGEN}) is satisfied with
\es{hijGEN}{
h_{ij} ( \bold{g}) = \delta_{ij}  + {\cal O}(\bold{g})   \,,
}
so the matrix $h_{ij}({\bf g})$ is positive definite to first nonvanishing order in ${\bf g}$.  Of course, as long as the perturbative expansion converges, $h_{ij}({\bf g})$ will continue to be positive definite at the very least in a small neighborhood of ${\bf g} = 0$.  A potential route towards proving the $F$-theorem is to construct the metric $h_{ij} ( \bold{g})$ non-perturbatively and demonstrate that it is positive definite.  Such an approach was undertaken in \cite{Friedan:2003yc} for one-dimensional field theories that can be realized as boundaries of two-dimensional field theories.

\section{$F$-coefficients for free conformal fields}

\subsection{Free conformal scalar field}

In this section we calculate the free energy of a free scalar field conformally coupled to the round $S^d$.  Similar results have appeared in \cite{QuineChoi, Kumagai}.

In $d$-dimensions the action of a free scalar field conformally coupled to $S^d$ is given by
 \es{ActionScalar0}{
  S_S =  \frac{1}{2} \int d^d x\, \sqrt{G} \left[ \left( \nabla \phi \right)^2 + \frac {d-2}{4(d-1)} R \phi^2  \right] \,.
 }
We take the radius of the round $S^d$ to be $a$, so that the Ricci scalar is $R = d(d-1)/a^2$.  Up to a constant additive term,
 \es{FFormal}{
  F_S = -\log | Z_S |=  \frac{1}{2} \log \det \left[ \mu_0^{-2} {\cal O}_S \right] \,,  \qquad {\cal O}_S \equiv -\nabla^2 + \frac {d-2}{4(d-1)} R  \,.
 }
 where $\mu_0$ is the UV cutoff needed to properly define the path integral.  At the end of the day $F_S$ will not depend on $\mu_0$ or $a$ in odd dimensions.  When $d \geq 2$, the eigenvalues of ${\cal O}_S$ are
 \es{OBEvalues}{
  \lambda_n = \frac{1}{a^2} \left(n + \frac{d}{2} \right) \left(n - 1 + \frac d2 \right)\,,   \qquad  n \geq 0 \,,
 }
and each has multiplicity
 \es{multiplicity}{
  m_n = \frac{(2n + d - 1) (n+d-2)!}{(d-1)! n!} \,.
 }
The free energy is therefore
 \es{FEvaluesGeneral}{
  F_S = \frac 12 \sum_{n=0}^\infty m_n \left[ -2
   \log  (\mu_0 a) + \log \left(n + \frac{d}{2} \right) + \log \left(n - 1 + \frac d2 \right) \right] \,.
 }
This sum clearly diverges at large $n$, but it can be regulated using zeta-function regularization.  By explicit computation, one can see that, unlike in even dimensions, in odd dimensions we have
 \es{Summ}{
  \sum_{n=0}^\infty m_n = 0 \, ,
 }
so there is no logarithmic dependence on $\mu_0 a$.  This is in agreement with the fact that there is no conformal anomaly in this case.  The remaining contribution to this sum can be computed from the function
 \es{NewFunction}{
  -\frac{1}{2} \sum_{n=0}^\infty \left[ \frac{m_n}{\left(n + \frac{d}{2} \right)^s} +  \frac{m_n}{\left(n -1 + \frac{d}{2} \right)^s} \right] =  -\frac{1}{2} \sum_{n=0}^\infty \frac{m_n + m_{n-1}}{\left(n -1+ \frac{d}{2} \right)^s} \,,
 }
whose derivative at $s = 0$ formally gives \eqref{FEvaluesGeneral}.  One can check that $m_n + m_{n-1}$ is a polynomial of degree $d-1$ in $n -1+ \frac{d}{2}$, so the sum in \eqref{NewFunction} converges absolutely for $s>d$ and can be evaluated in terms of $\zeta(s - k, \frac{d}{2} - 1)$, with $k$ ranging over all even integers between $0$ and $d-1$.  In $d=3$, for example, we have $m_n = (n+1)^2$ and $m_n + m_{n-1} = 2 \left(n+\frac 12\right)^2 + \frac 12$, so
\es{BosonRewriteSum}{
  F_S = - \frac{1}{2} \frac{d}{ds} \left[2 \zeta \left(s-2, \frac 12\right) + \frac 12 \zeta\left(s, \frac 12\right)\right] \bigg \vert_{s=0}
   = \frac{1}{16} \left( 2 \log 2 - \frac{3 \zeta(3)}{ \pi^2} \right) \approx 0.0638\,.
   }
For other values of $d$, see table~\ref{FBTable}.   These results agree with earlier work  \cite{QuineChoi, Kumagai, Marino:2011nm}.  For all odd $d$ we note that the the free energy on $S^d$ equals minus the entanglement entropy across $S^{d-2}$ calculated in \cite{Dowker:2010yj}.

\begin{table}
\begin{center}
\begin{tabular}{c|l}
 $d$ & $F_S$ \\
 \hline
 $3$ & $\frac{1}{2^4} \left(2 \log 2 - \frac{3 \zeta(3)}{\pi^2} \right) \approx 0.0638$ \\
 $5$ & $\frac{-1}{2^8}\left(2 \log 2 + \frac{2 \zeta(3)}{\pi^2} -
    \frac{15 \zeta(5)}{\pi^4}\right) \approx -5.74 \times 10^{-3}$ \\
 $7$ & $\frac{1}{2^{12}} \left( 4 \log 2 + \frac{82 \zeta(3)}{15 \pi^2} - \frac{10 \zeta(5)}{\pi^4}
    - \frac{63 \zeta(7)}{\pi^6} \right) \approx 7.97 \times 10^{-4}$ \\
 $9$ & $\frac{-1}{2^{16}} \left( 10 \log 2 + \frac{1588 \zeta(3)}{105 \pi^2}
    - \frac{2 \zeta(5)}{\pi^4} - \frac{126 \zeta(7)}{\pi^6} - \frac{255 \zeta(9)}{\pi^8} \right)
    \approx -1.31 \times 10^{-4}$ \\
 $11$ & $\frac{1}{2^{20}} \left( 28 \log 2 + \frac{7794 \zeta(3)}{175 \pi^2} + \frac{1940 \zeta(5)}{63 \pi^4}
    - \frac{1218 \zeta(7)}{5 \pi^6} - \frac{850 \zeta(9)}{\pi^8} - \frac{1023 \zeta(11)}{\pi^{10}}  \right)
    \approx 2.37 \times 10^{-5}$
\end{tabular}
\caption{The $F$-coefficient for a free conformal scalar field on $S^d$.}
\label{FBTable}
\end{center}
\end{table}

\subsection{Free massless fermion field}

In this section we calculate the free energy of a  free massless complex Dirac fermion on the round $S^d$.  We begin with the free fermion action
 \es{FermionAction}{
  S_D = \int d^d x\, \sqrt{G} \psi^\dagger (i \slashed{D}) \psi  \, .
 }
 Unlike in the case of the free conformal scalar action, the conformal fermion action does not contain a coupling between the fermion fields and curvature.  The free energy is given by
  \es{FreeFermion}{
  F_D = -\log | Z_D| = - \log \det \left[\mu_0^{-1} {\cal O}_D \right]
     \,, \qquad {\cal O}_D \equiv i \slashed{D}  \, .
 }
 The eigenvalues of ${\cal O}_D$ are
 \es{OFEvalues}{
  \pm \frac{1}{a} \left(n + \frac d2 \right)  \,, \qquad n \geq 0\,,
 }
each with multiplicity
 \es{MultFermions}{
  \hat m_n = \dim \gamma \begin{pmatrix}
   n + d -1 \\
   n
  \end{pmatrix} \,.
 }
Here, $\dim \gamma$ is the dimension of the gamma matrices in $d$ dimensions.   For odd $d$, $\dim \gamma = 2^{\frac{d-1}{2}}$ in the fundamental representation.

One can then write \eqref{FreeFermion} as
 \es{FFermionsEvalues}{
  F_D = -2 \sum_{n = 0}^\infty \hat m_n \left[- \log (\mu_0 a) + \log \left(n + \frac d2 \right) \right] \, ,
 }
and again one can check that $\sum_{n=0}^\infty \hat m_n = 0$ in odd dimensions using zeta-function regularization, so there is no logarithmic dependence on $\mu_0 a$.  To compute $F_D$, one can write it formally as the derivative at $s=0$ of the function
 \es{FFFunction}{
  2 \sum_{n=0}^\infty \frac{\hat m_n}{\left(n + \frac d2 \right)^s} \, .
 }
One can check that $\hat m_n$ is a polynomial of degree $d-1$ in $n + \frac d2$, so the sum in \eqref{FFFunction} converges absolutely for $s>d$ and can be expressed in terms of $\zeta(s-k, \frac d2)$, with $k$ ranging over the even integers between $0$ and $d-1$.

For $d = 3$, $\hat m_n = (n+2)(n+1)$, and the $F$-coefficient of a massless Dirac fermion is
 \es{FZetas}{
  F_D = 2 \zeta'(-2, 3/2) - \frac 12 \zeta'(0, 3/2)
   = \frac{\log 2}{4} + \frac{3\zeta(3)}{8 \pi^2} \approx 0.219 \,.
 }
 This result agrees with earlier work  \cite{Marino:2011nm}. \footnote{We note that $F_D/F_S$ is not a rational number and is quite large, $\approx 3.43$. For comparison, we note that the contribution of a $d=3$ massless Dirac fermion to the thermal free energy is $3/2$ times that of a massless real scalar. In $d=4$ the $a$-coefficient of a massless Dirac fermion is $11$ times that of a conformal scalar,
 while its contribution to the thermal free energy is $7/2$ times that of a massless scalar.
 Only in $d=2$ does the $c$-coefficient of a massless Dirac fermion equal that of a massless scalar.}
  For other values of $d$, see table~\ref{FFTable}.  The $F$-coefficient of a Majorana fermion is one half the result in table~\ref{FFTable}.

\begin{table}
\begin{center}
\begin{tabular}{c|l}
 $d$ & $F_D / \dim \gamma$ \\
 \hline
 $3$ & $\frac{1}{2^4} \left(2 \log 2 + \frac{3 \zeta(3)}{\pi^2} \right) \approx 0.110$ \\
 $5$ & $\frac{-1}{2^8}\left(6 \log 2 + \frac{10 \zeta(3)}{\pi^2} +
    \frac{15 \zeta(5)}{\pi^4}\right) \approx -2.16 \times 10^{-2}$ \\
 $7$ & $\frac{1}{2^{12}} \left( 20 \log 2 + \frac{518 \zeta(3)}{15 \pi^2} + \frac{70 \zeta(5)}{\pi^4}
    + \frac{63 \zeta(7)}{\pi^6} \right) \approx 4.61 \times 10^{-3}$ \\
 $9$ & $\frac{-1}{2^{16}} \left( 70 \log 2 + \frac{12916 \zeta(3)}{105 \pi^2}
    + \frac{282 \zeta(5)}{\pi^4} + \frac{378 \zeta(7)}{\pi^6} + \frac{255 \zeta(9)}{\pi^8} \right)
    \approx -1.02 \times 10^{-3} $ \\
 $11$ & $\frac{1}{2^{20}} \left( 252 \log 2 + \frac{234938 \zeta(3)}{525 \pi^2} + \frac{69124 \zeta(5)}{63 \pi^4}
    + \frac{8778 \zeta(7)}{5 \pi^6} + \frac{1870 \zeta(9)}{\pi^8} + \frac{1023 \zeta(11)}{\pi^{10}}  \right)
    \approx 2.32 \times 10^{-4}$
\end{tabular}
\caption{The $F$-coefficient for a free massless Dirac fermion field on $S^d$.  Here, $\dim \gamma$ is the dimension of the gamma matrices on $S^d$ and is equal to $2^{\frac{d-1}{2}}$ in odd dimensions $d$.}
\label{FFTable}
\end{center}
\end{table}

\subsection{Chern-Simons Theory}


In three dimensions $U(N)$ Yang-Mills theory does not have a UV fixed point. Instead, we will consider $U(N)$ Chern-Simons gauge theory with level $k$.  The $F$-coefficient for $N=1$ is $\frac{1}{2} \log k$, while for $N>1$ it was found to be \cite{Witten:1988hf}
 \es{FCS}{
  F_\text{CS}(k, N)= \frac{N}{2} \log (k+N) - \sum_{j=1}^{N-1} (N-j) \log \left (2 \sin \frac{\pi j}{k+N} \right )
   \,.
 }
In the weak coupling limit $k\gg N$, and for sufficiently large $N$, this expression may be approximated by
$ \frac{1}{2} N^2 \left (\log \frac {k} {2\pi N} + \frac {3}{2} \right)$.  Thus, somewhat surprisingly, the CS theory has a large $F$-coefficient, even though it has no propagating degrees of freedom.

In four dimensions, one of the first tests of the $a$-theorem was provided by the $SU(N)$ gauge theory coupled to $N_f$ massless Dirac fermions in the fundamental representation \cite{Cardy:1988cwa}.  This theory is asymptotically free for $N_f< 11 N/2$.   If this is the case, then in the UV the $a$-coefficient receives contributions from the $N_c^2-1$ gauge bosons and the $N_f N$ free fermions.  In the IR, it is believed that chiral symmetry breaking produces $N_f^2 - 1$ Goldstone bosons, which are the only degrees of freedom that contribute to $a_\text{IR}$. The asymptotic freedom condition $N_f< 11 N/2$ imposes an upper bound on the IR value of $a$ that is restrictive enough to not violate the $a$-theorem \cite{Cardy:1988cwa}.

In three dimensions we cannot construct similar tests involving $U(N)$ Yang-Mills theory coupled to fundamental fermions because the UV theory is not conformal. Instead we consider Chern-Simons gauge theories.  As a first example take the $U(1)$ Chern-Simons gauge theory coupled to $N_f$ massless Dirac fermions of charge 1. For $k\gg 1$ this theory is weakly coupled, so the $F$-coefficient is
\es{Fuv}{
F_\text{UV}\approx \frac{1}{2} \log k + N_f \left ( \frac{\log 2}{4} + \frac{3\zeta(3)}{8 \pi^2} \right ) + O(N_f/k)\, .
}
Now, let us add a mass for the fermion. The IR fixed point is then described by the $U(1)$ Chern-Simons gauge theory with CS level $k\pm N_f/2$ generated through the parity anomaly \cite{Niemi:1983rq,Redlich:1983dv}, where the sign is determined by the sign of the fermion mass.
Therefore, the IR free energy is $\frac{1}{2} \log (k \pm N_f/2)$. It is not hard to check that this is smaller than \eqref{Fuv}.
 
 Now we consider $U(N)_k$ Chern-Simons gauge theory, $N>1$, coupled to $N_f$ massless fundamental Dirac fermions. For $k\gg N$ this is a weakly coupled conformal field theory whose
$F$-coefficient is
\es{FuvN}{
F_\text{UV} \approx \frac{1}{2} N^2 \left (\log \frac {k} {2\pi N} + \frac {3}{2} \right) + N N_f \left ( \frac{\log 2}{4} + \frac{3\zeta(3)}{8 \pi^2} \right ) +O(N_f N^2/k)\, .
}
Now, let us add a $U(N_f)$ symmetric mass for the fermions.
The IR fixed point is then described by the $U(N)$ Chern-Simons gauge theory with CS level $k\pm N_f/2$.
 Therefore, $F_\text{IR} = F_\text{CS}(k\pm N_f/2, N)$.
It is not hard to check that $F_\text{UV} > F_\text{IR}$ for any $N_f$ if $k\gg N$. The comparison is the simplest if, in addition, we assume $k\gg N_f$. Then
\es{Fir}{F_\text{IR} \approx \frac{1}{2} N^2 \left (\log \frac {k} {2\pi N} + \frac {3}{2} \right) \pm \frac{N_f N^2}{4k}+\ldots\ ,}
making it obvious that $F_\text{UV} > F_\text{IR}$.

\section{Double trace deformations}

In this section we study the change in free energy under a relevant double trace deformation in a $d$-dimensional large $N$ field theory, starting from a UV fixed point and flowing to an IR fixed point. Some of this section is a review of the earlier work
\cite{Gubser:2002vv,Diaz:2007an,Allais:2010qq}.

\subsection{Bosonic double trace deformation}
\label{bosdeff}

   Consider a bosonic single trace operator $\Phi$ within the UV conformal field theory.  Let the dimension of this operator, $\Delta$, lie inside the range $(d/2 - 1,d/2)$. The lower limit on the dimension is the unitarity bound.  We impose the upper limit on the dimension because we will be adding the operator $\Phi^2$ to the lagrangian and we want this to be a relevant operator.  There are general arguments \cite{Witten:2001ua,Gubser:2002vv} that this deformation will cause an RG flow to an IR fixed point where $\Phi$ has dimension $d - \Delta$.

We begin with the partition function
\es{doublepart}{
Z = \int D \phi \exp\left( -S_0 - \frac{\lambda_0}{2} \int d^dx \sqrt{G} \Phi^2 \right) \, = Z_0 \left\langle \exp\left( - \frac{\lambda_0}{2} \int d^dx \sqrt{G} \Phi^2 \right)\right \rangle_0 \,,
}
where, as in section~\ref{pert}, $\lambda_0$ is the bare coupling defined at the UV scale $\mu_0$, $\Phi$ is the bare operator, and expectation values $\langle \cdots \rangle_0$ are taken with respect to the conformal action $S_0$.  The measure $D \phi$ is schematic for integration over all degrees of freedom in the theory.  We are interested in calculating the difference $\delta F_\Delta$ between the free energies of the IR and UV fixed points,
\es{deltaF}{
\delta F_\Delta = - \log\left|\frac{Z}{Z_0}\right| \, .
}
We explicitly write $\delta F_\Delta$ as a function of the UV scaling dimension $\Delta$ to emphasize the dependence of the IR free energy on the UV scaling dimension of the single trace operator $\Phi$.

 As in~\cite{Gubser:2002vv} we proceed through a Hubbard-Stratonovich transformation.  That is, we introduce an auxiliary field $\sigma$ so that
\es{auxZ}{
\frac{Z}{Z_0} = \frac{1}{\int D \sigma \exp(\frac{1}{2 \lambda_0} \int d^dx \sqrt{G} \sigma^2)} \int D \sigma \left< \exp\left[ \int d^dx \sqrt{G} \left( \frac{1}{2\lambda_0} \sigma^2 + \sigma \Phi \right)\right] \right>_0 \,.
}
In this context, large $N$ implies that the higher point functions of $\Phi$ are suppressed relative to the two-point function by factors of $1/N$, where we take $N$ large.  This allows us to write
\es{expapprox}{
\left< \exp\left( \int d^dx \sqrt{G} \sigma(x) \Phi(x) \right) \right>_0 = \exp\left[ \frac{1}{2} \left< \left(\int  d^dx \sqrt{G} \sigma(x) \Phi(x) \right)^2 \right>_0 + o(1/N^0) \right] \,.
}
The integral in equation~\eqref{auxZ} is then simply a gaussian integral, which integrates to give
 \es{kZ}{
\delta F_\Delta = \frac{1}{2} \tr \log(K) \,,
 }
 where
 \es{k}{
 K(x,y) = \frac{1}{\sqrt{G(x)}} \delta(x-y) + \lambda_0 a^{d}  \left< \Phi(x) \Phi(y)\right>_0 \, .
 }

 We choose to normalize the operator $\Phi$ so that the perturbing operator $\Phi^2$ has the same normalization as the operator $O$ in section~\ref{pert}.  Specifically we take the two-point function of $\Phi$ on the round $S^d$ to be given by
 \es{twoptPhi}{
 \langle \Phi(x) \Phi(y) \rangle_0 = \frac{1}{\sqrt{2}} \frac{1}{s(x,y)^{2\Delta}} \, .
 }
  We then proceed by expanding the right hand side of equation~\eqref{twoptPhi} in $S^d$ spherical harmonics using
  \es{ylm}{
  \frac{1}{s(x,y)^{2\Delta}} = \frac{1}{a^{2\Delta}}\sum_{n,m} g_n Y_{n m}^*(x) Y_{n m}(y) \,,
  }
where we normalize the $Y_{n m}(x)$ to be orthonormal with respect to the standard inner product on the unit $S^d$.  The $g_n$ coefficients in equation~\eqref{ylm} can be found in~\cite{Gubser:2002vv}, where they are shown to be
\es{gl}{
g_n = \pi^{d/2} 2^{d-\Delta} \frac{\Gamma(\frac{d}{2} - \Delta)}{\Gamma(\Delta)} \frac{ \Gamma(n+\Delta)}{\Gamma(d+n-\Delta)} \,, \qquad n \geq 0 \, .
}

 The expression for $\delta F_\Delta$ \eqref{kZ} was evaluated using dimensional regularization in~\cite{Diaz:2007an} .  Here we briefly review their argument.  The eigenvalues of the operator $K$ only depend on the angular momentum $n$ through the $g_n$ coefficients of equation~\eqref{gl}.  States on the sphere $S^d$ with angular momentum $n$ have  the degeneracy $m_n$ given in equation~\eqref{multiplicity}.  One can therefore write the change in free energy as
 \es{deltafsum}{
\delta F_\Delta =  \frac{1}{2} \sum_{n=0}^{\infty} m_n \log\left[1+ \lambda_0 a^{d-2\Delta} g_n\right] \, .
 }
Because $d - 2 \Delta > 0$, in the IR limit $a^{d-2\Delta}$ goes to infinity.  Continuing to dimension $d<0$, the sum in equation~\eqref{deltafsum} converges and $\delta F_\Delta$ becomes
  \es{deltafsum1}{
\delta F_\Delta =  \frac{1}{2} \sum_{n=0}^{\infty} m_n \log\left(\frac{ \Gamma(n+\Delta)}{\Gamma(d+n-\Delta)}\right) \, ,
 }
where in simplifying equation~\eqref{deltafsum} one uses $\sum_n m_n = 0$ as in eq.~\eqref{Summ}.

The sum in equation~\eqref{deltafsum1} is evaluated exactly in~\cite{Diaz:2007an}.  In odd dimensions they find
\es{deltafsumodd}{
 \frac{d \left(\delta F_\Delta \right)}{d \Delta} = \frac{ (-1)^{(d+1)/2} \pi^2 (d- 2 \Delta)}{2 \Gamma(1+d)} \frac{ \sec\left[\pi \left( \Delta - \frac d 2 \right) \right] \tan\left[\pi \left( \Delta - \frac d 2 \right) \right] }{\Gamma(1-\Delta) \Gamma(1-d+\Delta)} \, .
 }
The result agrees exactly with the dual calculation in $AdS_{d+1}$ \cite{Diaz:2007an}.
  In the case of most interest, where $d=3$, it reduces to the following simple expression:
 \es{deltafsum3}{
  \frac{d \left(\delta F_\Delta \right)}{d \Delta} = -\frac{\pi}{6} (\Delta - 1) (\Delta - \frac{3}{2}) (\Delta - 2)  \cot (\pi \Delta) \,.
  }
As a last step, we integrate eq.~\eqref{deltafsum3} with respect to $\Delta$ to get the final expression for $\delta F_\Delta$,
  \es{deltafFinal}{
  \delta F_\Delta =  -\frac{\pi}{6} \int_{\Delta}^{3/2}  dx (x - 1) (x - \frac{3}{2}) (x - 2)  \cot (\pi x)  \, .
  }
 The upper limit of integration in eq.~\eqref{deltafFinal} is chosen to be $3/2$ because we know that $\delta F_{\Delta =3/2} =0$, as can be seen directly in eq.~\eqref{deltafsum1}, where each term in the sum vanishes when $\Delta = d/2$.  The reason why $\delta F_{\Delta =3/2} =0$ is that when $\Delta = d/2$ the operator $\Phi^2$ is marginal.

\begin{figure}[htb]
\begin{center}
\leavevmode
\scalebox{1}{\includegraphics{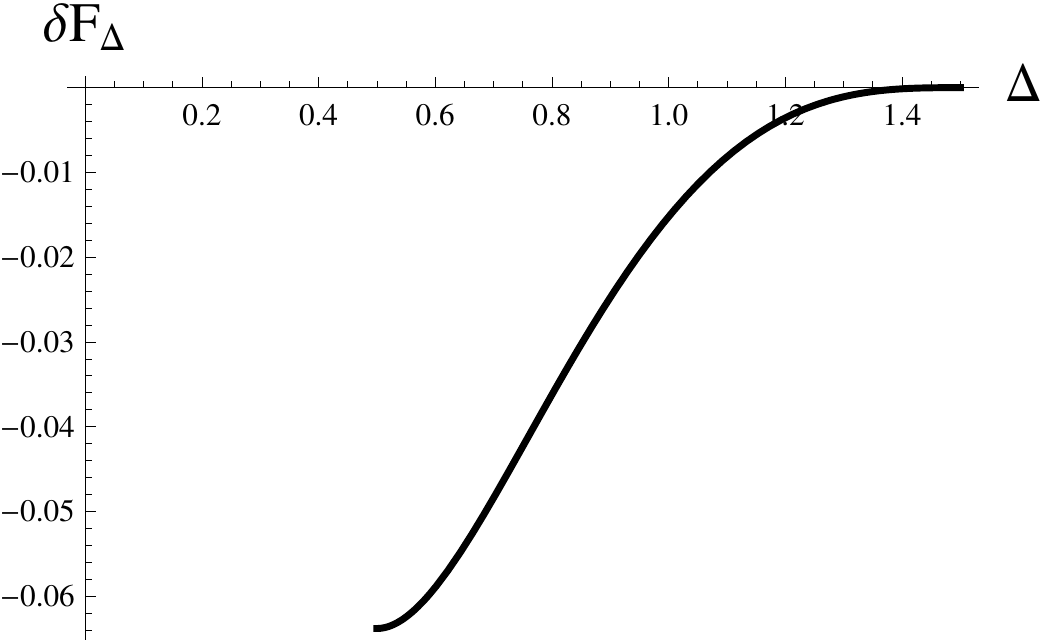}}
\end{center}
\caption{The change in free energy $\delta F_\Delta =F_\text{IR}- F_\text{UV}$ when the UV theory is perturbed by a relevant double trace operator $O^2$, where $O$ has dimension $\Delta$.   }
\label{DELTAF}
\end{figure}

  In figure~\ref{DELTAF} we plot $  \delta F_\Delta $ over the complete range of $\Delta$ when $d = 3$.
There are two cases of special interest.  The first is when $\Delta =1$, which corresponds to the $O(N)$ models we discuss in section \ref{ON}.  The numerical value for the difference in the free energy between the IR and UV fixed points in this case is
\es{deltaf1}{
\delta F_{\Delta = 1} = - \frac{\zeta (3)}{8 \pi^2}  \approx -0.0152 \,.
}
The second case of interest is when $\Delta = 1/2$, so that the operator $\Phi^2$ corresponds to adding a mass term for the free scalar field $\Phi$.  In this case $\delta F $ evaluates to
\es{deltaf32}{
\delta F_{\Delta = 1/2} = -\frac{1}{16} \left(2 \log 2 - \frac{3 \zeta (3)}{\pi^2} \right) \approx -0.0638 \,.
}
The change in free energy in equation~\eqref{deltaf32} is simply minus the free energy of a massive real scalar field in equation~\eqref{BosonRewriteSum}.  This makes sense since in this case we simply integrated out the real free scalar field $\Phi$.  This result can be thought of as a check of our procedure.

There is one further consistency check we can easily perform.  If we take $\Delta = (3- \epsilon)/2 $, then the IR fixed point is the perturbative fixed point of section~\ref{pert}.  The coefficient of the three point function $C$ is easily calculated to be $C =  4/ \sqrt{2}$ in this case.  Equation~\eqref{deltafdiff} predicts that the difference in free energy between the IR and UV fixed points is
\es{deltaFpertcheck}{
\delta F_{\Delta = (3-\epsilon)/2}= -\frac{\pi^2 \epsilon^3}{576} + o(\epsilon^3) \, .
}
Indeed, expanding the integral in equation~\eqref{deltafFinal} for  $\Delta = (3 - \epsilon)/2$ with $\epsilon$ small reproduces exactly equation~\eqref{deltaFpertcheck}.  This provides another consistency check between the double trace calculation and the perturbative calculation.

\subsection{RG flows in $O(N)$ vector models}
\label{ON}
In this section we discuss RG flows in the $O(N)$ vector models and compare the free energies of the various fixed points.  We begin with the classical $O(N)$ model action in flat $3$-dimensional Euclidean space,
\es{ONlag}{
S[\vec{\Phi}] = \frac{1}{2} \int d^3x \left[ \partial \vec{\Phi} \cdot \partial \vec{\Phi} + m_0^2 \vec{\Phi}^2 + \frac{\lambda_0}{2N} \left(\vec{\Phi} \cdot \vec{\Phi} \right)^2 \right] \,,
}
where $\vec{\Phi}$ is an $N$-component vector of real scalar fields.  The $F$-coefficient of the UV fixed point of this theory is of course just that of $N$ massless free scalar fields: $F_\text{UV}^\text{bos}= \frac{N}{16} \left(2 \log 2 - 3 \frac{\zeta(3)}{\pi^2} \right)$. If we take $m^2 > 0$, then all $N$ scalar fields become massive in the IR and we end up with the trivial empty theory whose $F$-coefficient vanishes.  The critical model comes from maintaining the vanishing renormalized mass.  This theory has a non-trivial IR fixed point.  The difference between the IR and UV $F$-coefficients is given in equation~\eqref{deltaf1}.  Therefore for large $N$ the $F$-coefficient of the IR fixed point in the critical $O(N)$ model is
\es{fcrit}{
F_{\text{crit}}^\text{bos} =\frac{N}{16} \left(2 \log 2 - 3 \frac{\zeta(3)}{\pi^2} \right) - \frac{\zeta(3)}{8 \pi^2} + O(1/N).
}
The free and critical $O(N)$ vector models have been conjectured \cite{Klebanov:2002ja} to be dual to the minimal Vasiliev higher-spin gauge theory in $AdS_4$ \cite{Vasiliev:1992av,Vasiliev:1995dn,Vasiliev:1999ba},
with different boundary conditions. Recently, this conjecture has been subjected to some non-trivial tests
\cite{Giombi:2009wh,Giombi:2010vg,Giombi:2011ya}, and new ideas have appeared on how to prove it \cite{Koch:2010cy,Douglas:2010rc}. It would be very interesting to match
our field theory results for $F_\text{UV}^\text{bos}$ and $F_\text{crit}^\text{bos}$ using the higher spin theory in Euclidean $AdS_4$.

Now consider perturbing the critical $O(N)$ model by the scalar mass term with $m_0^2 < 0$.  As the theory flows to the IR the potential breaks the symmetry from $O(N)$ to $O(N-1)$, and so by Goldstone's theorem we pick up $N-1$ flat directions in field space.  In the far IR these Goldstone modes simply become $N-1$ free massless scalar fields, with $F$-coefficient
\es{fcritAgain}{
F_{\text{Goldstone}} =\frac{N-1}{16} \left(2 \log 2 - 3 \frac{\zeta(3)}{\pi^2} \right) \ .
}
Thus,
\es{critGold}{F_{\text{Goldstone}} - F_{\text{crit}}^\text{bos}= - \frac{1}{16} \left(2 \log 2 - 5 \frac{\zeta(3)}{\pi^2} \right)
\approx -0.0486}
in agreement with the conjectured $F$-theorem.

This conclusion should be contrasted with the evolution of the thermal free energy coefficient $c_{\text{Therm}}$. In the critical $O(N)$
model $c_{\text{Therm}} = 4N/5 + O(1)$ \cite{Sachdev:1993pr,Chubukov:1994zz}, while in the Goldstone phase $c_{\text{Therm}} = N-1$.
Thus, for large enough $N$ the flow from the critical $O(N)$ model to the Goldstone phase rules out the possibility of a $c_{\text{Therm}}$ theorem.  On the other hand, the coefficient of the stress-energy tensor 2-point function $c_T$ decreases when the $O(N)$ model flows from the critical to the Goldstone phase \cite{Petkou:1995vu}.  Thus, such a flow does not rule out the possibility of a $c_T$ theorem.

Another interesting $O(N)$ model to consider is the $d=3$ Gross-Neveu model with $N$ massless Majorana fermions $\psi^i$ and the
interaction term $(\bar \psi^i \psi^i)^2$. This model has an interacting UV fixed point where the pseudoscalar operator $ \bar \psi^i \psi^i$ has dimension $1 + O(1/N)$. The IR fixed point is described simply by $N$ free fermions.
Thus, we find that
\es{Fferm}{
F_\text{UV}^\text{ferm} &=\frac{N}{16} \left(2 \log 2 + 3 \frac{\zeta(3)}{\pi^2} \right) + \frac{\zeta(3)}{8 \pi^2} + O(1/N) \ ,\\
F_\text{IR}^\text{ferm} &=\frac{N}{16} \left(2 \log 2 + 3 \frac{\zeta(3)}{\pi^2} \right)\ .
}
The higher-spin duals of these theories in $AdS_4$ were conjectured in \cite{Leigh:2003gk,Sezgin:2003pt}, and recently these conjectures were subjected to non-trivial tests \cite{Giombi:2009wh,Giombi:2010vg}. It would be interesting to derive the results (\ref{Fferm}) using the higher-spin gauge theory in Euclidean $AdS_4$.

\subsection{Fermionic double trace deformation}
\label{fermdouble}

In this section we study the change in free energy under a fermionic double trace deformation in a large $N$ field theory on $S^d$.  The calculation proceeds analogously to that in section~\ref{bosdeff}, where we deformed the UV fixed point by a bosonic double trace deformation.  The difference is that we replace the bosonic operator $\Phi(x)$ by a fermionic, single-trace operator $\chi (x)$.  In this section we will assume that $\chi$ is a complex Grassmann-valued spinor field.  In order to obtain the difference in free energy between the IR and UV fixed points when $\chi$ is Majorana, all one has to do is divide the final result by two.

Let the dimension of the operator $\chi$ be $\Delta$, with $\Delta$ inside the range $[(d - 1)/2,d/2]$.  The lower limit on the dimension is the unitarity bound on spinor operators.  The upper limit on the dimension comes from requiring the operator $\bar{\chi} \chi$ to be relevant.  Just as in the case of the bosonic double trace deformation, one can argue that the double-trace deformation will induce an RG flow that takes the theory to an IR fixed point where $\chi$ has dimension $d - \Delta$ \cite{Allais:2010qq}.

We want to compute the $F$-coefficient of the IR fixed point, so we need to calculate the free energy of the theory on the round $S^d$.  The partition function on $S^d$ is given by
\es{doublepartAgain}{
Z = Z_0 \left< \exp\left( - \lambda_0 \int d^dx \sqrt{G} \bar{\chi} \chi \right)\right>_0 \,,
}
where $\lambda_0$ is the coupling of dimension $d-2\Delta$.  The calculation of the expectation value in equation~\eqref{doublepartAgain} was presented in~\cite{Allais:2010qq}, and here we summarize their derivation.  First, we introduce a complex auxiliary spinor field $\eta$ and write
\es{auxZFerm}{
\frac{Z}{Z_0} = \frac{1}{\int D \eta D \bar{\eta} \exp(\int d^dx \sqrt{G} \bar{\eta} \eta)} \int D \eta D \bar{\eta}\left< \exp \left[ \int d^dx \sqrt{G} \left( \bar{\eta} \eta + \sqrt{\lambda_0} ( \bar{\eta} \chi + \bar{\chi} \eta) \right) \right] \right>_0  \,.
}
Just as in the bosonic case, the assumption of large $N$ comes into play by taking the expectation value inside of the exponential, giving
\es{fermapprox}{
\left< \exp \left[ \int d^dx \sqrt{G} \sqrt{\lambda_0} ( \bar{\eta} \chi + \bar{\chi} \eta) \right] \right>_0 =  \exp \left[ \lambda_0 \int d^dx \sqrt{G} \int d^dy \sqrt{G} \bar{\eta}(x) \left< \chi(x) \bar{\chi}(y) \right>_0 \eta(y) + o(1/N^0) \right]\,.
}
We assume that $n$-point functions, with $n>2$, are suppressed by inverse power of $N$.  The integral in equation~\eqref{auxZFerm} is then Gaussian.  Exponentiating the result to give the change in free energy $\delta F_\Delta$ we find
\es{deltaFFerm}{
\delta F_\Delta = - \tr \log(\hat{K}) \,,
}
where
 \es{khat}{
 \hat{K}(x,y) = \frac{1}{\sqrt{G(x)}}\delta(x-y) + \lambda_0 a^d \left<\chi(x) \bar{\chi} (y)\right>_0 \, .
 }

In flat space we choose the fermion two-point function to have the normalization
\es{twopointFerm}{
\hat{G}(x,y) = \left<\chi(x) \bar{\chi} (y)\right>_0 = \frac{ \gamma \cdot (x-y)}{\abs{x-y}^{2\Delta+1}} \,.
}
We need to find the eigenvalues and degeneracies of the operator $\hat{G}$ on the sphere.  This problem is solved in~\cite{Allais:2010qq} and here we simply quote the result.\footnote{See section~5.3 of that paper.}    The eigenvalues
\es{eigenFerm}{
\hat{g}_n  \propto \pm i \frac{\Gamma(n+\Delta+1/2)}{\Gamma(n+d-\Delta + 1/2)} \, , \qquad n \geq 0
}
come in conjugate pairs and are indexed by the integer $n$ that runs from zero to infinity.  In equation~\eqref{eigenFerm} we leave off any $n$ independent proportionality factors, because as we will see below these factors do not contribute to the free energy in the IR limit.   At each level $n$ there is a degeneracy $\hat m_n$ given in equation~\eqref{MultFermions}.

\begin{figure}[htb]
\begin{center}
\leavevmode
\scalebox{1}{\includegraphics{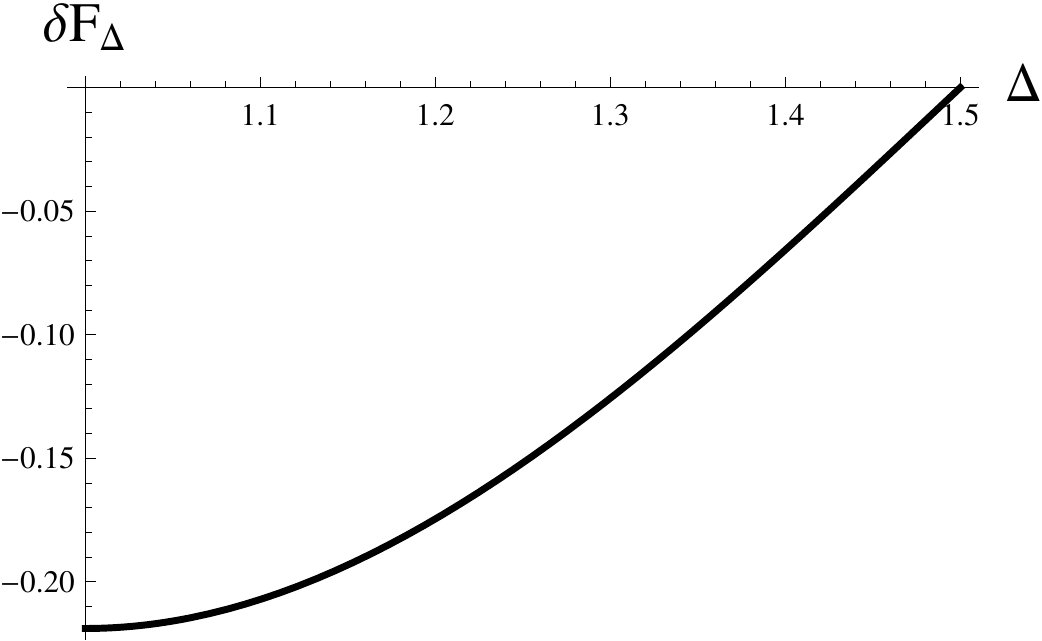}}
\end{center}
\caption{The change in free energy $\delta F_\Delta = F_\text{IR}- F_\text{UV}$ when the UV theory is perturbed by a relevant double trace operator $\bar{\chi} \chi$, where $\chi$ is a fermionic operator of dimension $\Delta$.   }
\label{DELTAFFERM}
\end{figure}

  Analytically continuing to the region of the complex plane where $\Re(d) < 1$, the trace in equation~\eqref{deltaFFerm} converges and in the IR limit we can write
\es{deltaFerm}{
\delta F_\Delta = - 2 \sum_{n=0}^{\infty} \hat m_n \log{\frac{\Gamma(n+\Delta+1/2)}{\Gamma(n+d-\Delta + 1/2)}} \,.
}
The sum in equation~\eqref{deltaFerm} is easily evaluated using the methods in~\cite{Diaz:2007an}.  After a simple calculation one finds the result (specifying to three-dimensions)
\es{deltaFFermint}{
\delta F_\Delta = -\frac{2\pi}{3} \int_{\Delta}^{3/2} dx \left( x - \frac{1}{2} \right) \left( x - \frac{3}{2} \right) \left( x - \frac{5}{2} \right) \tan (\pi x) \, , \qquad \Delta \in \left(1,\frac 3 2\right) \, .
}
In figure~\ref{DELTAFFERM} we plot the change in free energy $\delta F$ over this range of $\Delta$.

Just as in the bosonic case, we can check this procedure by evaluating $\delta F$ when $\chi$ has the dimensionality $\Delta = 1$ of a free spinor field.  The integral in equation~\eqref{deltaFFermint} evaluates to
\es{DeltaFFermFree}{
\delta F_{\Delta = 1} = - \frac{1}{8} \left( 2 \log 2 +  3\frac{\zeta(3)}{\pi^2} \right) \approx -.219 \,.
}
Comparing to equation~\eqref{FZetas}, we see that this is the $F$-coefficient of a massless complex spinor field.  Intuitively this makes sense, because in this case all we have done is to integrate out a massive free complex spinor.

We note that, as in section 5.1, $\delta F$ vanishes for $\Delta=3/2$ where the double-trace operator is marginal.
One might be tempted to expand equation~\eqref{deltaFFermint} for $\Delta$ near $3/2$ and attempt to compare with the perturbative result in equation~\eqref{deltafdiff}.  One would quickly find that this does not work.  Indeed, letting $\Delta =( 3 - \epsilon)/2$ we see that
\es{fermFexpand}{
\delta F_{\Delta = (3 - \epsilon)/2} = -\frac{\epsilon}{3} + O(\epsilon^3) \, .
}
The leading change in the free energy is order $\epsilon$ while in the perturbative calculation of section~\ref{pert} the change in free energy is order $\epsilon^3$.

The resolution is that the perturbing operator $O = \bar \chi \chi$ is a pseudo-scalar, as can be seen, for example, from the fact that the correlation functions of an odd number of $O(x_i)$ change sign under $x_i \to - x_i$.  From eq.~\eqref{twopointFerm}, one can compute explicitly the connected correlation functions in flat space:
\es{threepointferm}{
\langle O(x) O(y) \rangle &= \frac{\dim (\gamma)}{\abs{x-y}^{4\Delta}} \,, \\
\langle O(x)O(y) O(z) \rangle_0 &=  2 i \dim (\gamma) \frac{ \epsilon_{i j k} (x-y)_i (y - z)_j (z-x)_k}{( | x-y| | y-z| |z-x| )^{2 \Delta+1}} \,, \\
\langle O(x) O(y) O(z) O(w) \rangle_0 &= -\dim (\gamma) \biggl(2 X_{xzyw} X_{xwyz} + 2X_{xyzw} X_{xzyw}
  + 2X_{xyzw} X_{xwyz} \\
  {}&+ X_{xyyz} X_{xwzw} + X_{xyxw}X_{yzzw} \biggr) \,,
}
where we have defined
 \es{XDef}{
  X_{abcd} \equiv \frac{(a-b) \cdot (c-d)}{\abs{a-b}^{2 \Delta + 1} \abs{c-d}^{2 \Delta + 1}} \,.
 }
That the $n$-point functions change sign under reflection implies that only odd powers of the coupling appear in the beta function, and also that only even powers of the coupling appear in an expansion of the free energy as in eq.~\eqref{ffcn}.   We believe that all coefficients in the expansions of the beta function and $\delta F$ as power series in the coupling constant are ${\cal O}(\epsilon)$, as can be checked explicitly for the case of the four-point function in~\eqref{threepointferm}.   It follows that the IR fixed point does not occur when the coupling $g$ is small.  One would therefore need to calculate all the terms in eq.~\eqref{ffcn} in order to find the change in the $F$-coefficient along the RG flow.


\section*{Acknowledgments}

We thank A.~Dymarsky, S.~Giombi, C.~Herzog, D.~Jafferis, J.~Maldacena, S.~Sachdev, E.~Witten, and X.~Yin for useful discussions. IRK thanks the Simons Center for Geometry and Physics at Stony Brook (in particular, the 9th annual Simons Summer Workshop in Mathematics and Phyiscs), the Center for the Fundamental Laws of Nature at Harvard University, IHES, and the DAU Project in Kharkov for hospitality and opportunities to present talks on this work. This research was supported in part by the US NSF under Grant No.~PHY-0756966. IRK gratefully acknowledges support from the IBM Einstein Fellowship at the IAS and from the John Simon Guggenheim Memorial Fellowship. The work of SSP was also supported in part by a Pappalardo Fellowship in Physics at MIT, and by Princeton University through a Porter Ogden Jacobus Fellowship.  The work of BRS was supported in part by the NSF Graduate Research Fellowship Program.


\appendix

\section{Comments on Massive Free Fields}

At conformal fixed points we can map a $d$-dimensional theory in flat, odd-dimensional space to the Euclidean $S^d$ and compute the unique $F$-coefficient.  However, away from conformal fixed points the mapping is ambiguous.  In this Appendix we search for a function that interpolates between conformal fixed points by looking at the example of massive free fields in three dimensions.

\subsection{Free massive scalar field}
\label{mboson}

We consider the action
 \es{ActionScalar1}{
  S_S =  \frac{1}{2} \int d^3 x\, \sqrt{G} \left[ \left( \nabla \phi \right)^2 + \frac 18 R \phi^2 + m^2 \phi^2 \right]
 }
 on $S^3$.  This leads to the following infinite sum for the free energy
  \es{FEvalues1}{
  F_S = \frac{1}{2} \sum_{n= 1}^\infty n^2 \log  \left[ n^2 - \frac 14 + (a m)^2 \right] \,,
 }
 which clearly diverges as $n \to \infty$.  The dimensionless parameter $(a m)$ flows from zero at the UV conformal fixed point to infinity in the IR.  For notational convenience we set $a=1$, so that the RG scale is simply given by the mass $m$.   We calculate
 \es{derF}{
   \frac{\partial F_S}{\partial (m^2)} = \frac{1}{2} \sum_{n=1}^\infty \frac{n^2}{n^2 + m^2 - \frac 14}
   = \frac{1}{2} \left( \sum_{n=1}^\infty 1 - \sum_{n=1}^\infty \frac{ m^2 - \frac 14 }{n^2 + m^2 - \frac 14}\right) \,.
 }
Using zeta-function regularization, $\sum_{n=1}^\infty 1 = \zeta(0) = -1/2$.  The resulting sum is convergent and gives
 \es{derFAgain}{
  \frac{\partial F_S}{\partial (m^2)} = -\frac 14 \pi \sqrt{m^2 - \frac 14} \coth \left[ \pi \sqrt{m^2 - \frac 14} \right] \,.
 }
  While the function in equation~\eqref{derFAgain} is manifestly negative and vanishes when $m=0$, it asymptotes to a linear function of $m$ as $m \to \infty$. Thus, with this regularization we do not find the desired result
  \es{intdfdmb}{
 \int_{0}^\infty d m^2 \frac{ \partial F_S}{\partial m^2} = - \frac{1}{16} \left( 2 \log 2 - \frac{3 \zeta(3)}{ \pi^2} \right) \,.
 }

 The problem seems to be that we have generated a finite cosmological constant of order $m^3$ and a finite coefficient
 of the $\sqrt{G} R$ term of order $m$ along the flow.
We will therefore modify the free energy by
  \es{deltaSB}{
  \delta F_S = f[( am )] \, ,
  }
  where $f[(a m )]$ is some function which must vanish at $ am=0$ and cancels the undesirable terms at large $am$.  Again we take $a=1$ so that the change in the expression for the free energy is simply $\delta F_S = f(m)$.

  We can calculate the necessary form of $f(m)$ at the IR fixed point.    Suppose the IR fixed point were at $m = \Lambda$, where $\Lambda$ is assumed to be large.  We require $ \partial F_S ( m = \Lambda) / \partial m^2= 0$, which implies that
  \es{deltaFB}{
\left. \frac{ \partial f}{\partial m^2} \right|_{m=\Lambda}  &= \frac{\pi}{4} \sqrt{\Lambda^2 - \frac 1 4} \coth\left[\pi \sqrt{ \Lambda^2 - \frac{1}{4}} \right] \\
 &= \frac{\pi \Lambda}{4} - \frac{\pi}{32 \Lambda} + O(1/ \Lambda^3)  \, ,
}
in the limit of large $\Lambda$.  We then infer that in the IR the function $f(m)$ takes the form
\es{deltaFB1}{
f(m) = \frac{\pi m^3}{6} - \frac{\pi m}{16} + O(1/m) \,.
}
 Two examples of functions which obey \eqref{deltaFB1} as well as $f(0) = 0$ are
 \es{deltaFB2}{
 f_1(m) =  \frac{\pi m^3}{6} - \frac{\pi m}{16} \, , \qquad  f_2(m) = \frac{\pi}{6} \left( m^2 - \frac 1 4 \right)^{3 \over 2} \theta\left(m^2 - \frac 1 4\right) .
 }

One finds that with the function $f_1(m)$ the free energy is not monotonically decreasing and that $\partial F_S / \partial m^2$ diverges at $m=0$.  The function $f_2(m)$ has the advantage that with it the free energy is a monotonically decreasing function of $m$.  It is not an analytic function of the mass, though.  Interestingly, one finds by an explicit calculation that with either $f_1(m)$ or $f_2(m)$ we find the desired result
 (\ref{intdfdmb}).

\subsection{Free massive fermion field}
\label{ferm}

In this section we perform an analogous computation to that in section~\ref{mboson} but for a massive complex fermion field.  Naively the massive fermion action on $S^3$ is given by
 \es{FermionAction1}{
  S_D = \int d^3 x\, \sqrt{G} \left[\psi^\dagger (i\slashed{D}) \psi - i m \psi^\dagger \psi \right] \,.
 }
By a now familiar computation we can write the free energy as
 \es{FreeFermionAgain1}{
  F_D = -\sum_{n=1}^\infty n(n+1) \log \left[ \left(n+\frac 12 \right)^2 + m^2 \right] \,.
 }
 Taking a derivative of equation~\eqref{FreeFermionAgain1} with respect to $m^2$ and performing zeta-function regularization we arrive at
  \es{derFFermionAgain}{
  \frac{\partial F_D}{\partial (m^2)} = \frac{4 m^2 + 1}{8 m} \pi \tanh (\pi m) \,.
 }

 Like in the case of the massive scalar field, we see that at large $m$ the right hand side of equation~\eqref{derFFermionAgain} asymptotes to a linear function of $m$, which signals a cosmological constant of order $m^3$.  We again add a function $\delta F_D = f(m)$ to the free energy so that it has the correct asymptotic form as $m \to \infty$.  Explicitly we take
 \es{fmferm}{
 f(m) = - \frac{\pi \abs{m}}{3} \left(m^2 + \frac 3 4\right) \, .
 }
With this correction the function $F_D(m)$ is monotonically decreasing between the UV and IR fixed points and by an explicit computation one can check
 \es{intdfdmf}{
 \int_{0}^\infty d m^2 \frac{ \partial F_D}{\partial m^2} = - \frac{\log 2}{4} - \frac{3\zeta(3)}{8 \pi^2} \,,
 }
which is the desired result.

\bibliographystyle{ssg}
\bibliography{ThreeSphere}

\end{document}